\documentclass[final,authoryear,5p,times,twocolumn]{elsarticle}

\usepackage{graphicx}
\graphicspath{{../Figs/}}

\usepackage{amssymb}

\usepackage[switch]{lineno}

\usepackage{amsmath}
\usepackage{bm}
\usepackage{hyperref}
\usepackage{color}
\usepackage{threeparttable}
\usepackage{array}
\usepackage{multirow}
\usepackage{url}
\usepackage[english]{babel}
\usepackage{fixltx2e}
\usepackage{afterpage} 

\journal{Medical Image Analysis}

\newcommand{\PreserveBackslash}[1]{\let\temp=\\#1\let\\=\temp}
\newcolumntype{C}[1]{>{\PreserveBackslash\centering}p{#1}}
\newcolumntype{R}[1]{>{\PreserveBackslash\raggedleft}p{#1}}
\newcolumntype{L}[1]{>{\PreserveBackslash\raggedright}p{#1}}

\begin{document}
\begin{frontmatter}



\title{Three-Dimensional Segmentation of the Left Ventricle in Late Gadolinium Enhanced MR Images of Chronic Infarction Combining Long- and Short-Axis Information}


\author[ECE]{Dong Wei\corref{cor1}}
\ead{dongwei@nus.edu.sg}

\author[ECE]{Ying Sun}
\ead{elesuny@nus.edu.sg}

\author[ECE]{Sim-Heng Ong}
\ead{eleongsh@nus.edu.sg}

\author[NUHS]{Ping Chai}
\ead{ping{\_}chai@nuhs.edu.sg}

\author[NUH]{Lynette L. Teo}
\ead{lynette{\_}ls{\_}teo@nuhs.edu.sg}

\author[NUHS]{Adrian F. Low}
\ead{adrian{\_}low@nuhs.edu.sg}

\address[ECE]{Department of Electrical and Computer Engineering, National University of Singapore, Singapore 117576, Republic of Singapore}

\address[NUHS]{Cardiac Department, National University Heart Centre, 5 Lower Kent Ridge Road, Singapore 119074, Republic of Singapore}

\address[NUH]{Department of Diagnostic Imaging, National University Hospital, 5 Lower Kent Ridge Road, Singapore 119074, Republic of Singapore}

\cortext[cor1]{Corresponding author. Tel.: +65 65168035, fax: +65 67773117. }

\begin{abstract}
Automatic segmentation of the left ventricle (LV) in late gadolinium enhanced (LGE) cardiac MR (CMR) images is difficult due to the intensity heterogeneity arising from accumulation of contrast agent in infarcted myocardium.
In this paper, we present a comprehensive framework for automatic 3D segmentation of the LV in LGE CMR images.
Given myocardial contours in cine images as \emph{a priori} knowledge, the framework initially propagates the \emph{a priori} segmentation from cine to LGE images via 2D translational registration.
Two meshes representing respectively endocardial and epicardial surfaces are then constructed with the propagated contours.
After construction, the two meshes are deformed towards the myocardial edge points detected in both short-axis and long-axis LGE images in a unified 3D coordinate system.
Taking into account the intensity characteristics of the LV in LGE images, we propose a novel parametric model of the LV for consistent myocardial edge points detection regardless of pathological status of the myocardium (infarcted or healthy) and of the type of the LGE images (short-axis or long-axis).
We have evaluated the proposed framework with 21 sets of real patient and 4 sets of simulated phantom data.
Both distance- and region-based performance metrics confirm the observation that the framework can generate accurate and reliable results for myocardial segmentation of LGE images.
We have also tested the robustness of the framework with respect to varied \emph{a priori} segmentation in both practical and simulated settings.
Experimental results show that the proposed framework can greatly compensate variations in the given \emph{a priori} knowledge and consistently produce accurate segmentations.
\end{abstract}


\end{frontmatter}


\section{Introduction}
\label{sec:Introduction}
Viability assessment of the myocardium after myocardial infarction due to ischemia is essential for diagnosis and therapy planning.
In particular, the detection, localization and quantification of the infarcted myocardium, also called infarct / infarction / scar, are important for determining whether and which part(s) of the myocardium may benefit from re-vascularization therapy.
Among various acquisition protocols used in cardiac magnetic resonance (CMR)\footnote{List of the abbreviations used in this paper (in numeric-alphabetic order): 2C: 2-chamber; 4C: 4-chamber; BP: blood pool; CMR: cardiac magnetic resonance; ECG: electrocardiography; LA: long-axis; LGE: late gadolinium enhanced; LV: left ventricle; MSSD: mean of sum of squared differences; MVO: microvascular obstruction; NCC: normalized cross correlation; NMI: normalized mutual information; PI: pattern intensity; ROI: region of interest; SA: short-axis; SSD: sum of squared differences.}
imaging, late gadolinium enhanced (LGE) imaging protocol offers the capability to directly visualize infarcts.
In a typical LGE CMR examination, a gadolinium-based contrast agent is injected and a single-frame sequence is acquired 10-20 minutes later, by which time the infarcts will exhibit hyper-enhanced intensities compared to healthy myocardium due to delayed wash-out kinetics of the contrast agent.
One exception is the no-reflow phenomenon called \emph{microvascular obstruction} (MVO), which is mostly observed in acute infarctions~\citep{MVO2010}.
In such cases, the involved sub-endocardial regions appear as dark as normal myocardium because no contrast agent can flow into these regions.
Figure~\ref{fig:LGE_example4Introduction} shows an LGE image with a hyper-enhanced infarct and MVO.

\afterpage{
\begin{figure}[tp]
    \centering
    \includegraphics[width=.49\textwidth]{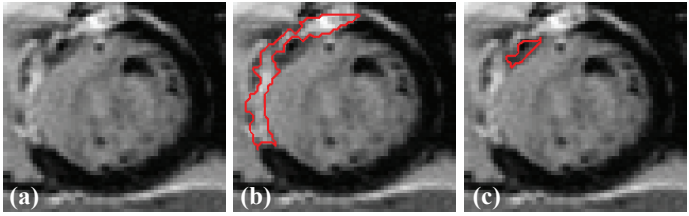}
    \caption{An example LGE image: (a) the original image; (b) the hyper-enhanced infarct is outlined. (c) the MVO is outlined.}\label{fig:LGE_example4Introduction}
\end{figure}
}

The delineation of myocardial contours is a prerequisite to automatic localization and quantification of infarcts in LGE images -- nearly all such works~\citep{infarctSeg2005,A_Comprehensive_Approach,infarctSeg2010a,infarctSeg2010b,infarctSeg2011} in the literature assume that high quality myocardial segmentation is given, either manually or (semi-) automatically.
Because manual delineation is not only time-consuming but also subject to inter-observer variability, it is highly desirable to automate the process.
However, the automation is difficult, due mainly to the intensity heterogeneity of the myocardium arising from the accumulation of the contrast agent in infarcted areas.

To the best of our knowledge, there has been little research aimed at fully automatic myocardial segmentation in LGE images.
Most existing approaches utilize pre-delineated myocardial contours in cine CMR data of the same patient as \emph{a priori} knowledge~\citep{relatedworks01_Ciofolo2008,relatedworks02_Dikici2004,myownwork}.
Such an approach is reasonable because the cine data, as the most frequently analyzed CMR data, is always acquired for a patient who is asked to remain still during the entire acquisition process.
This approach has two important advantages.
First, the pre-delineated LV in the cine data provides \textit{a priori} knowledge about the shape and appearance of the left ventricle (LV) in the LGE data, making the automatic segmentation more reliable, especially when the contrast between the blood pool (BP) and infarcted myocardium is poor.
Second, many methods have been proposed for (semi-)automatic segmentation of the cine data~\citep[][to name a few]{cineSegmentation_chen2008,cineSegmentation_hautvast2006,cineSegmentation_lichao2009}, and hence the entire process can be automated.
Nevertheless, major difficulties in this approach include:
(i) displacement and nonrigid deformation between cine and LGE data due to respiratory motion and inaccurate electrocardiography (ECG) gating; and
(ii) differences in resolution, field of view, and intensity characteristics of cine and LGE data.

In one of the pioneering works on automatic segmentation of LGE images,
\citet{relatedworks02_Dikici2004} proposed to obtain \emph{a priori} segmentation for the target LGE image by nonrigid registration of two nearby cine frames as well as of corresponding cine and LGE images,
and then deform the \emph{a priori} segmentation with a five-parameter affine transformation by maximizing the probability of correct segmentation.
Although elastic deformation of the LV between the \emph{a priori} segmentation and its actual shape in the LGE image was partly compensated for by interpolating the deformation field between the two nearby cine frames,
residual deformation may still remain because:
(i) the ECG gating is inherently imperfect; and
(ii) the interpolation of the deformation field is linear, while the deformation of the heart is known to be nonrigid.
Consequently, a global affine transformation is inadequate to capture shape changes of the myocardium.

More recently, \citet{relatedworks01_Ciofolo2008} proposed to first initialize and deform 2D myocardial contours according to image evidence in LGE images.
The myocardium was divided into four quadrants and those likely to contain large areas of infarcts were treated differently.
Subsequently, the 3D meshes that were pre-constructed from cine images were registered towards the stack of contours from LGE images.
It was novel to treat potential infarcts differently for the segmentation, but the division into four quadrants was too coarse to account for small infarcts.
In addition, there was no correction of misalignment artifacts.
\citet{heartMotionStudy} studied the motion of the heart due to respiration and found considerable displacements that should not be neglected.
For a 3D mesh representation of the LV, there is a high risk of the original physical shape being distorted if the misalignment of slices is not corrected.

In our previous work~\citep{myownwork}, we presented an automatic segmentation method that fully utilizes shared information between corresponding cine and LGE images.
Given myocardial contours in cine images, the segmentation of LGE images was achieved in a coarse-to-fine manner.
Affine registration was first performed between the corresponding cine and LGE image pair, followed by nonrigid registration, and finally local deformation of myocardial contours driven by forces derived from local features of the LGE image.
At the stage of local deformation, we proposed an adaptive detection of endocardial edges by selecting one of the two cases -- normal endocardium and sub-endocardial infarcts -- and also included an effective thickness constraint into the evolution scheme.
However, this method had the following drawbacks:
(i) short-axis (SA) images were segmented individually in 2D without utilizing the inherent 3D information;
(ii) the b-spline based nonrigid registration was slow; and
(iii) the adaptive detection of infarcted and healthy myocardial edges was primitive.

In this paper, we propose a novel 3D segmentation framework of the LV for LGE CMR images.
Our work is distinct from the aforementioned ones in three aspects.
(i) We integrate long-axis (LA) images (standard 4-chamber (4C) and 2-chamber (2C) views) with SA images for the 3D segmentation.
Besides providing complementary information about the LV between the largely spaced SA images, the LA images are also used for the correction of misalignment artifacts among slices.
(ii) We propose a novel parametric model of the LV for LGE images based on 1D intensity profiles.
This model is flexible to be applied to both SA and LA images, and self-adaptive to detect edge points of both infarcted and healthy myocardium.
Further, it detects paired endocardial and epicardial edge points simultaneously.
(iii) We introduce an effective thickness constraint for the 3D deformation scheme based on the simplex mesh geometry~\citep{simplexmeshIJCV1999}, which can also be applied to other scenarios involving coupled boundaries.

The rest of the paper is organized as follows.
Section~\ref{sec:Method} describes the proposed 3D segmentation framework.
Section~\ref{sec:Results} presents and discusses the experimental results on both real patient and simulated data.
Finally Section~\ref{sec:Conclusion} concludes the paper.

\section{Method}
\label{sec:Method}
Our 3D segmentation framework comprises four major steps:
(i) selection and pre-processing of target LGE images and corresponding cine images;
(ii) misalignment correction of the selected LGE images;
(iii) 2D translational registration to initially propagate the \emph{a priori} segmentation from cine to LGE data;
and (iv) 3D nonrigid deformation of the myocardial meshes (constructed from the propagated contours) driven by features in both SA and LA LGE images.

\subsection{Data selection and pre-processing}
\label{sec:Method:preparation}

First, we display all the SA LGE slices of one subject for user selection.
Typically, all the slices are chosen, except for the most basal slice(s) in which the myocardium is very thin and dim, and the most apical slice(s) in which the myocardium can hardly be discerned or the slices are already out of the LV.
The user also selects one 4C and one 2C LA LGE slice based on image contrast.
Once the target LGE slices are chosen, the corresponding SA cine images are automatically selected.
While a cine sequence includes multiple frames which cover the entire cardiac cycle, an LGE sequence has only one frame usually triggered between end-systole and end-diastole.
We select the cine image with the same slice location as and the closest phase\footnote{Due to the inherent imperfection of the ECG gating, exact phase matching is often impossible.}
to the LGE image according to the DICOM header information, and delineate the myocardial contours in the selected cine image as \emph{a priori} segmentation.
The delineation is done with a semi-automatic method~\citep{cineSegmentation_lichao2009} with manual corrections where necessary.
Finally, we normalize each pair of corresponding cine and LGE SA images to the same physical resolution (pixel size) by resizing the cine image.

\subsection{Misalignment correction}
\label{sec:Method:alignment}
\begin{figure}[tp]
  \center
  \includegraphics[width=0.45\textwidth]{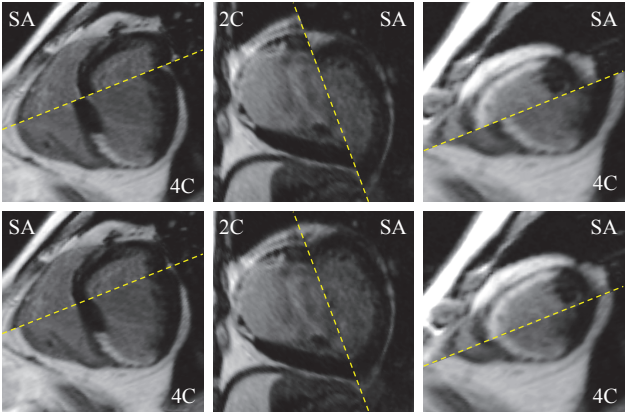}\\
  \caption{Misalignment correction results illustrated with intersections between the SA and LA slices. First row: original misaligned slices displayed according to the DICOM header information; second row: the same slices as in the first row, but after misalignment correction.}\label{fig:alignment}
\end{figure}

Misalignment artifacts between CMR slices is common due to one or more of the following causes:
(i) image acquisition during different breath holds;
(ii) image acquisition at different phases of different cardiac cycles; and
(iii) potential patient movement during the scan.
The first row of Fig.~\ref{fig:alignment} shows three examples with significant misalignments between the LA and SA slices.
Misalignment correction is indispensable to 3D reconstruction of both volume and surface;
otherwise, the reconstructed content would be distorted from its prototype.

Researchers have proposed various approaches to realign a set of CMR slices \citep{alignment3,alignment2,alignment1used,A_Comprehensive_Approach}.
Specially, \citet{alignment1used} treated cine images of the LV as planes in 3D space, and translated them by maximizing normalized mutual information \citep[NMI;][]{NMI1999} between intersecting line segments sampled on these planar images.
We implemented this method to realign LGE slices, but used the normalized mean of the sum of squared differences  (MSSD)\footnote{
That is, samples are normalized to zero mean and unit standard deviation before being used to compute MSSD.
} instead of NMI as the similarity measure.
Several realigned slices are shown in the second row of Fig.~\ref{fig:alignment}, demonstrating reasonable correction results.

\subsection{Translational registration}
\label{sec:Method:registration}
The segmentation starts by propagating the \emph{a priori} segmentation in the cine image to the LGE image by 2D translational registration.
In our previous work, we implemented a constrained affine registration using normalized cross correlation (NCC) as the similarity metric \citep{myownwork}.
However, after extensive experiments with different similarity metrics and degrees of freedom, we found that by using pattern intensity~\citep[PI;][]{PI_Introduction1997} as the similarity metric, a translational registration can already reach the same or sometimes better performance.

Given two images $I_{1}$ and $I_{2}$, PI operates on the difference image $I_{\mathrm{diff}}=I_{1}-I_{2}$.
If $I_{1}$ and $I_{2}$ are two well-registered images of the same object, structures from this object should vanish and there should be a minimum number of structures or patterns in $I_{\mathrm{diff}}$.
A suitable similarity measure should, therefore, characterize the \emph{structuredness} of $I_{\mathrm{diff}}$.
PI considers a pixel of $I_{\mathrm{diff}}$ to belong to a structure if it has a significantly different value from its neighboring pixels.
Using a constant radius $r$ to define the neighborhood, the PI is defined as:
\begin{linenomath}
\begin{equation}\label{eq:PI}
\begin{split}
  P_{r,\,\delta}&(I_{1},I_{2})=P_{r,\,\delta}(I_{\mathrm{diff}}) \\
    &=\frac{1}{N_{I_{\mathrm{diff}}}}\sum_{x,\,y}\frac{1}{N_{r}}\sum_{v,\,w}\frac{\delta^2}{\delta^2+[I_{\mathrm{diff}}(x,y)-I_{\mathrm{diff}}(v,w)]^2},
\end{split}
\end{equation}
\end{linenomath}
where $(x,y)$ denotes the pixels in $I_{\mathrm{diff}}$, and $(v,w)$ the neighboring pixels of $(x,y)$ within radius $r$;
they satisfy the relationship $(x-v)^2+(y-w)^2\leq r^2$.
$N_{I_{\mathrm{diff}}}$ and $N_{r}$ denote the number of pixels in $I_{\mathrm{diff}}$  and the neighborhood respectively.
A constant $\delta$ is introduced to suppress the impact of noise.

\begin{figure}[tp]
  \centering
  \includegraphics[width=0.475\textwidth]{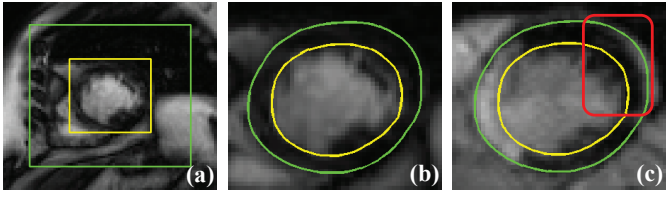}\\
  \caption{Illustration of the translational registration.
  (a) The cine image with bounding box of the LV (the yellow square) and the defined ROI (the green square) overlaid.
  (b) The cine image with pre-delineated contours overlaid.
  (c) The LGE image with translated contours overlaid. In general the contours segment the myocardium closely, but in the region highlighted with the red square, a discrepancy is observed.}\label{fig:rigid_translation}
\end{figure}

A rectangular region of interest (ROI) is defined in the cine image regarding the LV, and used as the matching window for the calculation of the similarity measure.
As shown in Fig.~\ref{fig:rigid_translation}(a), we first identify the bounding box of the LV in the cine image with the pre-delineated epicardial contour, and then define the ROI by enlarging the bounding box to twice its original size.
A window of the same size as the ROI is placed in the LGE image and translated to search for the best match, i.e., the one giving the maximum $P(I_{\mathrm{diff}})$ between the two windows.
After obtaining the optimal translation, the pre-segmented myocardial contours in the cine image (Fig.~\ref{fig:rigid_translation}b) are translated accordingly and become a coarse segmentation of the myocardium in the LGE image.
Figure~\ref{fig:rigid_translation}(c) shows an example of this translational registration.
In general the translated contours are close to the myocardial boundaries.
However, a discrepancy exists in the region highlighted with the red square.
Such discrepancies are often caused by the elastic deformation and out-of-plane motion of the heart, and different tissue structures such as pericardial fat.
Therefore, nonrigid deformation is needed to eliminate such discrepancies.

\subsection{Three-dimensional nonrigid deformation}
\label{sec:Method:deformation}
After translational registration, we apply a 3D nonrigid deformation to surface meshes constructed with the obtained coarse endocardial and epicardial contours (denoted by $\mathcal{C}_{\mathrm{endo,\,rigid}}$ and $\mathcal{C}_{\mathrm{epi,\,rigid}}$ afterwards).
First, we detect myocardial edge points in both SA and LA LGE images.
Second, we initialize two simplex meshes~\citep{simplexmeshIJCV1999} with $\mathcal{C}_{\mathrm{endo,\,rigid}}$ and $\mathcal{C}_{\mathrm{epi,\,rigid}}$, respectively.
Third, we deform the meshes with both external and internal forces.
The final meshes are naturally a 3D segmentation of the LV.

\subsubsection{A Novel Parametric Model of the LV in LGE Images}
\label{sec:Method:deformation:model}
Taking into account intensity characteristics of the myocardium in LGE images, we propose a novel parametric model of the LV based on 1D intensity profiles~\citep{1dProfile01,1dProfile02,1dProfile03}.
This model has the following desirable properties:
it automatically adapts to either healthy or infarcted myocardium and consistently detects reliable myocardial edge points;
it is flexible to be applied to both SA and LA images with minor alteration;
it detects paired endocardial and epicardial edge points at the same time with variable thickness of the myocardium;
and it does not make the assumption that enhancements always reside sub-endocardium in every infarcted SA image, which was made by our earlier work~\citep{myownwork}.

\begin{figure*}[tp]
  \centering
  \includegraphics[width=\textwidth]{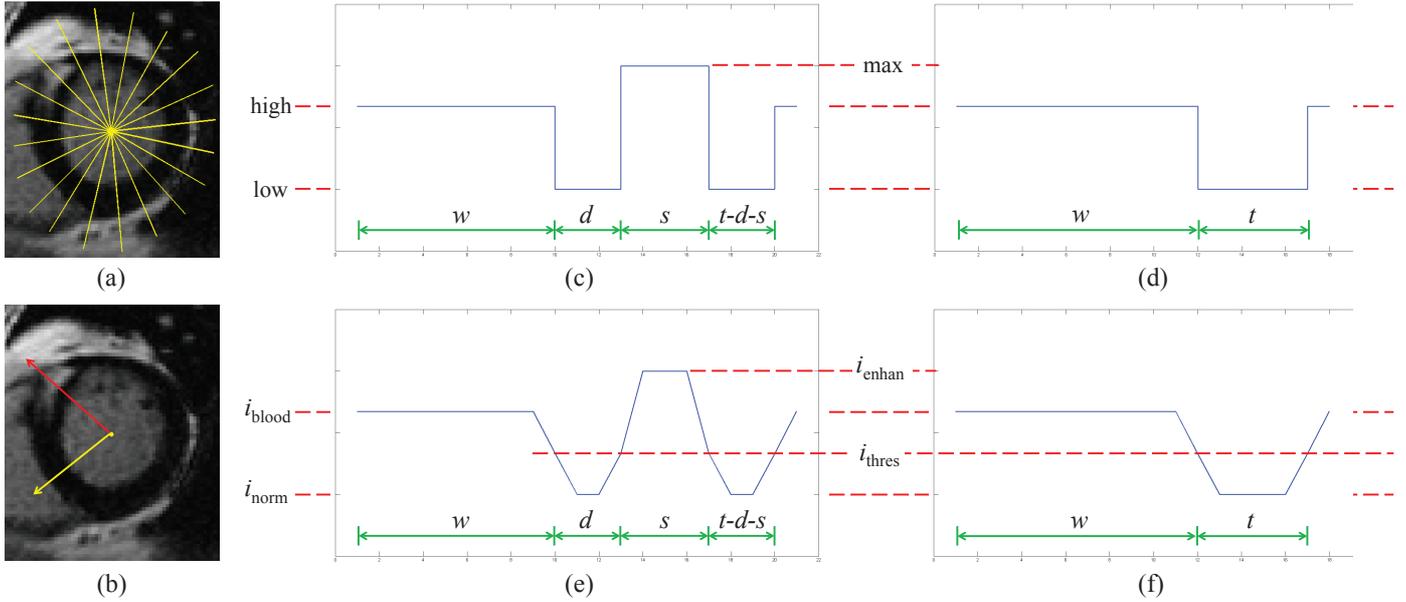}\\
  \caption{(a) A representative SA LGE image is sampled along evenly spaced rays.
  (b) Yellow ray: a sample ray corresponding to normal myocardium; red ray: a sample ray corresponding to infarcted myocardium.
  (c) An intensity profile template $I_{\mathrm{templt}}(w,\, t,\, s,\, d)$ devised to model the case of infarcted myocardium.
  (d) An intensity profile template simplified from (c), i.e., $s=0$, to model the case of normal myocardium.
  (e)-(f) More realistic intensity profile templates with values estimated from the LGE image and gradual transitions.
  Note: for better illustration, relative lengths of $w$, $t$, $s$ and $d$ in (c)-(f) do not strictly follow the two sample rays in (b).
  }\label{fig:rays_n_profiles}
\end{figure*}

We now introduce the proposed model with reference to an SA image (Fig.~\ref{fig:rays_n_profiles}).
We find the LV center $O_{\mathrm{LV}}$ by averaging all the contour points on $\mathcal{C}_{\mathrm{endo,\,rigid}}$ and $\mathcal{C}_{\mathrm{epi,\,rigid}}$, and then sample 1D intensity profiles along 79 evenly spaced radial directions emanating from $O_{\mathrm{LV}}$ (Fig.~\ref{fig:rays_n_profiles}(a)).
These intensity profile samples are denoted by $I_{\mathrm{sample}}(\theta)$.

We model 1D intensity patterns along the radial rays from $O_\mathrm{LV}$ to beyond the epicardium with intensity profile templates $I_\mathrm{templt}(w,t,s,d)$, where $w$ denotes the distance from $O_{\mathrm{LV}}$ to the endocardium, $t$ the thickness of the myocardium, $s$ the thickness of the enhancement and $d$ the distance from the endocardium to the enhancement.
The endocardial and epicardial edge points can be expressed by the parameters $w$ and $t$:
\begin{linenomath}
\begin{equation}\label{eq:wtpara}
\begin{split}
    &\mathcal{C}_{\mathrm{endo}}(\theta) = O_{\mathrm{LV}} + \bm{n}(\theta) \cdot w(\theta), \\
    &\mathcal{C}_{\mathrm{epi}}(\theta) = O_{\mathrm{LV}} + \bm{n}(\theta) \cdot [w(\theta) + t(\theta)],
\end{split}
\end{equation}
\end{linenomath}
where $\bm{n}$ denotes the unit vector along the emanating ray.
For every sampled angle $\theta$, if $w(\theta)$, $t(\theta)$, $s(\theta)$ and $d(\theta)$ are correct, they would produce the $I_{\mathrm{templt}}(w,\, t,\, s,\, d)$ which is most similar to $I_{\mathrm{sample}}(\theta)$
\footnote{
Because $I_{\mathrm{sample}}(\theta)$ is purposely sampled far beyond epicardium and thus always longer than the devised $I_{\mathrm{templt}}(w,\, t,\, s,\, d)$, when we compare the two only the first $w(\theta)+t(\theta)$ pixels of $I_{\mathrm{sample}}(\theta)$ are used.
}.
Therefore, the problem of finding $\mathcal{C}_{\mathrm{endo}}$ and $\mathcal{C}_{\mathrm{epi}}$ is now converted to searching for the desired $(w_\mathrm{d},\, t_\mathrm{d},\, s_\mathrm{d},\, d_\mathrm{d})$ for each $\theta$.
Let $Err(I_{1},\,I_{2})$ denote a mismatch measurement function, for which we adopt the commonly used MSSD.
Thus the desired $(w_\mathrm{d},\, t_\mathrm{d},\, s_\mathrm{d},\, d_\mathrm{d})$ is given by:
\begin{linenomath}
\begin{equation}\label{eq:optimalwts}
    (w_{\mathrm{d}},\, t_{\mathrm{d}},\, s_{\mathrm{d}},\, d_{\mathrm{d}}) = \arg \min\{Err[I_{\mathrm{templt}}(w,\, t,\, s,\, d),\, I_{\mathrm{sample}}]\}.
\end{equation}
\end{linenomath}

For infarcted myocardium (e.g., the red ray in Fig. \ref{fig:rays_n_profiles}(b)), an exemplary $I_\mathrm{templt}{(w,\, t,\, s,\, d)}$ is shown in Fig.~\ref{fig:rays_n_profiles}(c).
The first $w$ pixels are set to high (the BP), followed by $d$ pixels low (the un-enhanced myocardium), then $s$ pixels maximum (the enhancement should display a higher intensity than the BP in an \emph{ideal} case) and the remaining $(t-d-s)$ pixels low.
This template is able to cover most enhancement patterns found in chronic infarct patients of ischemic heart disease:
for sub-endocardial enhancements, $d=0$ and $s<t$;
for transmural enhancements, $d=0$ and $s=t$;
for mid-myocardium enhancements, such as in cases of MVO, $d>0$ and $0<s<t$.
For normal myocardium (e.g., the yellow ray in Fig.~\ref{fig:rays_n_profiles}~(b)), $s=0$ and the value of $d$ has no impact (but has to satisfy $0\leq d\leq t$), and the corresponding profile template is illustrated in Fig.~\ref{fig:rays_n_profiles}~(d).

\begin{figure}[tp]
  \centering
  \includegraphics[width=0.49\textwidth]{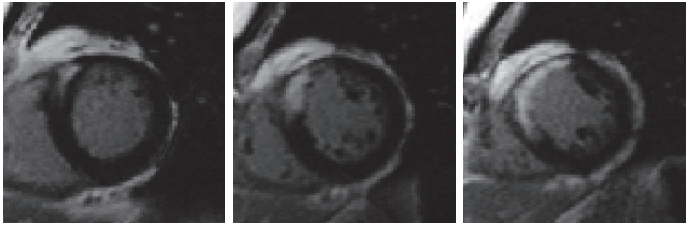}\\
  \caption{From left to right: three consecutive SA LGE slices. The infarcts grow from sub-endocardium in 3D. However, when considering the leftmost slice alone, the same conclusion can hardly be drawn due to the loss of 3D spatial continuity.}\label{fig:slice_stack}
\end{figure}

Since nearly all infarcts originate from the sub-endocardium \citep{medical_scar_origination01, medical_scar_origination02}, our previous work assumed that enhancements always reside there in every SA image with infarcts~\citep{myownwork}.
However, this assumption is occasionally violated in practice.
One such example is shown in Fig.~\ref{fig:slice_stack}, where three consecutive SA LGE slices are displayed.
The infarcts do grow from the sub-endocardium in 3D.
But when the leftmost slice is treated alone, the assumption no longer holds due to the loss of 3D spatial continuity.
A possible solution is to keep this assumption but impose it in 3D~\citep{assumptionIn3D}.
Alternatively, we completely relax the assumption and introduce the parameter $d(\theta)$ to describe the position of the enhancement within the myocardium.

Given the translated epicardial contours, representative intensity values are estimated from the LGE images and used to devise the intensity profile templates.
All the pixels enclosed by $\mathcal{C}_{\mathrm{epi,\,rigid}}$ in all the SA slices of the same subject are classified into two classes by Otsu's threshold~\citep{OtsuThreshold1979} $i_{\mathrm{thres}}$.
Then $i_{\mathrm{norm}}$, the mean of all the pixels below $i_{\mathrm{thres}}$ is used as the value of normal myocardium when devising the templates.
All the pixels above $i_{\mathrm{thres}}$ are further classified by k-means clustering into two clusters, whose centers ($i_{\mathrm{blood}}$ and $i_{\mathrm{enhan}}$) are used as the values of the BP and enhancements, respectively.
To mimic the gradual transition between bright and dark regions, $i_{\mathrm{thres}}$ is used as the transition value.
With these values, we can devise intensity profile templates which resemble the samples with higher fidelity (Fig.~\ref{fig:rays_n_profiles}~(e) and (f)).

\subsubsection{Myocardial edge points detection in SA images}
\label{sec:Method:deformation:edgePointsSA}
Although the parametric model is flexible and adaptive, sometimes it may be trapped by similar but false edges such as the edge between the thin slice of fat and the lung surrounding the LV due to its 1D nature.
In order to avoid such traps by imposing continuity constraints on individual 1D intensity profiles, we incorporate this model in an energy minimization scheme.
This scheme is applicable to both SA and LA images, but we first describe it using SA images.
The application of the parametric model and the energy minimization scheme to LA images will be described in the next section.

The energy to be minimized comprises an intensity profile match term and a smoothness term weighted by a constant $\lambda$:
\begin{linenomath}
\begin{equation}\label{eq:E_whole}
    E = E_{\mathrm{match}} + \lambda \cdot E_{\mathrm{smooth}}.
\end{equation}
\end{linenomath}
$E_{\mathrm{match}}$ is simply a summation of the mismatch measurements over all the sampled radial angles $\theta$:
\begin{linenomath}
\begin{equation}\label{eq:E_profile}
    E_{\mathrm{match}} = \sum_{\theta} Err[I_{\mathrm{templt}}(w(\theta),\, t(\theta),\, s(\theta),\, d(\theta)),\, I_{\mathrm{sample}}(\theta)].
\end{equation}
\end{linenomath}
Since both the endocardial and epicardial contours are parameterized as $\mathcal{C}(\theta)=(w(\theta),\, t(\theta))$, we define $E_{\mathrm{smooth}}$ as
\begin{linenomath}
\begin{equation}\label{eq:E_smooth}
    E_{\mathrm{smooth}} = \sum_{\theta}[ (\mathcal{C}_{\theta})^2 + (\mathcal{C}_{\theta\theta})^2 ].
\end{equation}
\end{linenomath}
By minimizing $E_{\mathrm{smooth}}$, the endocardial and epicardial contours tend to deform into two concentric circles.
The energy in (\ref{eq:E_whole}) is minimized with $w$ and $t$ constrained within a narrow band of $(w_{\mathrm{0}},\, t_{\mathrm{0}})$, which are the original $w$ and $t$ computed from $\mathcal{C}_{\mathrm{endo,\,rigid}}$ and $\mathcal{C}_{\mathrm{epi,\,rigid}}$.

Next, we discuss the influence of papillary muscles on the proposed model.
Because we do not consider the papillary muscles when devising $I_{\mathrm{templt}}$, their presence does influence the mismatch measurement $Err(I_{\mathrm{templt}},\, I_{\mathrm{sample}})$.
However, by incorporating the parametric model into the energy minimization scheme, the detection of myocardial edge points is made robust to the papillary muscles in two ways:
(i) by using a constrained optimization within a narrow band of the original $(w_{\mathrm{0}},\, t_{\mathrm{0}})$, the papillary muscles resulting in $w$'s very distant from $w_{\mathrm{0}}$ are directly eliminated;
(ii) the continuity constraints on both $w$ and $t$ can prevent endocardial edge points from being dragged too far from their neighbors by the papillary muscles.

\begin{figure}[tp]
  \centering
  \includegraphics[width=.475\textwidth]{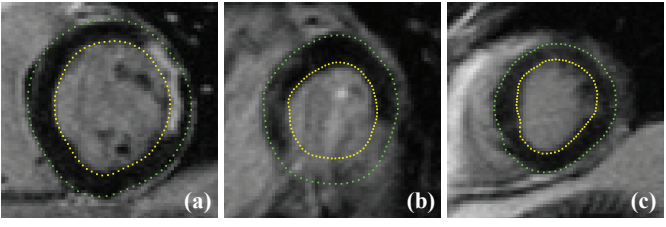}\\
  \caption{Examples of the detected myocardial edge points in SA images.
  Even when there is a large area of transmural infarcts completely blending in with the BP and surrounding tissues in (b), our method still provides reasonable myocardial edge points.}\label{fig:featurepts_SA}
\end{figure}

Figure~\ref{fig:featurepts_SA} shows three examples of the detected myocardial edge points in SA images.
Notably in Fig.~\ref{fig:featurepts_SA}~(b), where there is a large area of transmural infarcts completely blending in with the BP and surrounding tissues, our method still provides reasonable myocardial edge points because of the smoothness term.

\subsubsection{Myocardial edge points detection in LA images}
\label{sec:Method:deformation:LAfeatures}
\begin{figure*}[tp]
  \centering
  \includegraphics[width=.95\textwidth]{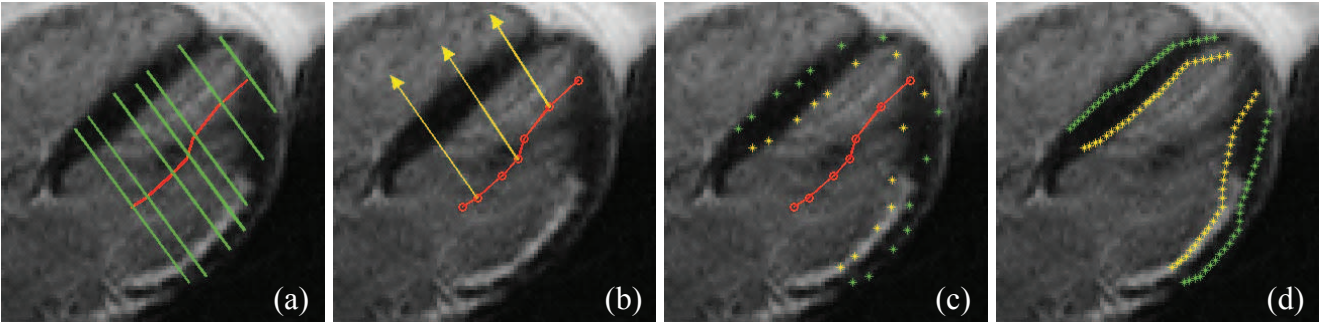}\\
  \caption{(a) SA slice locations are used to parameterize myocardial contours in LA images.
  (b) 1D intensity profiles $I_{\mathrm{sample}}$ are sampled along the rays pointing from the central axis of the LV to beyond the epicardium.
  (c) Intersection points of the LA image with $\mathcal{C}_{\mathrm{endo,\,rigid}}$ and $\mathcal{C}_{\mathrm{epi,\,rigid}}$; they are used as starting points of the search for myocardial edge points and to estimate the central axis of the LV.
  (d) More points are interpolated between SA slices, in order to fully utilize information contained in the LA images.}\label{fig:intensity_profile_model_LA}
\end{figure*}

Different from the parameterizing variable used for SA images (i.e., radial angles $\theta$), slice locations $l$ along the normal to the SA image planes are used to parameterize myocardial contours in LA images (Fig.~\ref{fig:intensity_profile_model_LA}~(a)).
Hence, myocardial contours are expressed as $\mathcal{C}(l)=(w(l),\, t(l))$.
\footnote{Note that in an LA image each $l$ actually defines dual sets of $\mathcal{C}_{\mathrm{endo}}(l)$ and $\mathcal{C}_{\mathrm{epi}}(l)$, as the myocardium is divided into two sides by the central axis of the LV.
Since we treat the two sides \emph{separately} in exactly the same way, we use only one side for the purpose of elaboration.}
We sample the LA images along rays pointing from the central axis of the LV to beyond the epicardium (Fig.~\ref{fig:intensity_profile_model_LA}~(b));
the samples are denoted as $I_{\mathrm{sample}}(l)$.
Again, we devise $I_{\mathrm{templt}}(w,t,s,d)$ with tentative $(w,t,s,d)$ and search for the optimal tetrad which makes $I_{\mathrm{templt}}$ most resemble $I_{\mathrm{sample}}$ for each $l$.

Now two questions are how to estimate:
(i) the central axis of the LV, and
(ii) reasonable starting points for the search of the optimal $(w,t,s,d)$.
To answer both questions, we intersect $\mathcal{C}_{\mathrm{endo,\,rigid}}$ and $\mathcal{C}_{\mathrm{epi,\,rigid}}$ with the LA image planes.
The intersection points serve as the initial coarse myocardial edge points in the LA images.
Points on the central axis can be easily estimated by averaging the four intersection points on the same SA slices (Fig.~\ref{fig:intensity_profile_model_LA}(c)).
The intersections also naturally define the parametrization scheme (i.e., values of $l$) by the relative location of each SA slice with respect to the LA image planes.
However, the SA slices are too sparse and thus there are too few $l$'s.
On one hand, the SA slices are far away from each other, causing the loss of anatomical continuity of the myocardium;
on the other hand, LA images actually contain \emph{far more} information than what is provided by the intersected locations.
In order to retain the natural smoothness of the myocardium and utilize as much information in LA images as possible, we make $l$ denser by linearly interpolating between neighboring SA slices (Fig.~\ref{fig:intensity_profile_model_LA}(d)).

\begin{figure}[tp]
  \centering
  \includegraphics[width=.49\textwidth]{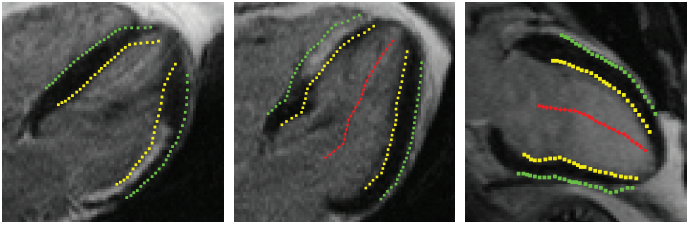}\\
  \caption{Examples of the detected myocardial edge points in LA images. The first two images are 4C views, while the last one is a 2C view.}\label{fig:featurepts_LA}
\end{figure}

We use the same energy minimization scheme for LA images as for SA images.
The energy functional to be minimized still comprises two terms $E_{\mathrm{match}}$ and $E_{\mathrm{smooth}}$ as in (\ref{eq:E_whole}).
$E_{\mathrm{match}}$ stays the same as in (\ref{eq:E_profile}), except that the parametrization variable is $l$.
However, $E_{\mathrm{smooth}}$ in (\ref{eq:E_smooth}) must be slightly adapted to accommodate the shape of the LV in LA images.
That is, in LA images $w$ tends to gradually shrink from the base to apex of the LV.
Therefore, the first order derivative of $w$ is removed from (\ref{eq:E_smooth}) and $E_{\mathrm{smooth}}$ becomes:
\begin{linenomath}
\begin{equation}\label{eq:E_smooth_LA}
    E_{\mathrm{smooth}} = \sum_{l}[ (w_{ll})^2 + (t_{l})^2 + (t_{ll})^2 ].
\end{equation}
\end{linenomath}
Several examples of the detected myocardial edge points in LA images are shown in Fig.~\ref{fig:featurepts_LA}.

\subsubsection{The deformation scheme}
\label{sec:Method:deformation:scheme}
Unlike the previous step in which the coarse segmentation is achieved by registration of corresponding cine and LGE images, in this step we directly deform the contours based on local features of the LGE image.
We use the \emph{simplex mesh} \citep{simplexmeshIJCV1999} geometry and a Newtonian law of motion to represent and deform the contours.
Both endocardial and epicardial surfaces are represented with a simplex mesh respectively, and deformed in an inter-related manner.
At each vertex we define three forces, namely, smoothness force $\bm{F}_{\mathrm{smooth}}$, edge attraction force $\bm{F}_{\mathrm{edge}}$ and myocardium thickness force $\bm{F}_{\mathrm{thick}}$.
$\bm{F}_{\mathrm{smooth}}$ imposes uniformity of vertex distribution and continuity of simplex angles~\citep{simplexmeshIJCV1999}.
$\bm{F}_{\mathrm{edge}}$ draws vertices towards detected myocardial edge points along the normal direction at each vertex.
Finally, $\bm{F}_{\mathrm{thick}}$ maintains proper distances between paired endocardial and epicardial vertices.
Letting $p^{t}$ denote the position of any vertex at time $t$, the deformation scheme is given by
\begin{linenomath}
\begin{equation}\label{eq:ffdScheme}
    p^{t+1}=p^{t}+(1-\gamma)(p^{t}-p^{t-1})+\alpha\bm{F}_{\mathrm{smooth}}+\beta\bm{F}_{\mathrm{edge}}+\mu\bm{F}_{\mathrm{thick}},
\end{equation}
\end{linenomath}
where $\gamma$ is a damping factor, and $\alpha$, $\beta$ and $\mu$ are respective weights.

We initialize the simplex mesh for the epicardium with $\mathcal{C}_{\mathrm{epi,\,rigid}}$ in the steps described below.

Above all, we transform all the involved coordinates to the image coordinate system of the last SA slice.
We then re-sample $\mathcal{C}_{\mathrm{epi,\,rigid}}$ with evenly spaced radial angles.
The resampling makes the connection relationship among mesh vertices straightforward by aligning all vertices on all slices along the same radial angles, hence simplifying mesh construction.
After that we interpolate to obtain several planar contours evenly spaced between the physically existing SA slices.
The interpolation is necessary because
(i) considering the large distance between neighboring SA slices, properly densified vertices can make $\bm{F}_{\mathrm{smooth}}$ more meaningful;
and (ii) some of the interpolated vertices will be attracted by myocardial edge points detected in LA images, which in turn can affect the vertices on the physically existing SA slices through $\bm{F}_{\mathrm{smooth}}$.

Now we specify the connection relationship among vertices.
We use the vertices on the two boundary SA slices (the most basal and apical ones) to form two \emph{planar 1-simplex meshes} \citep{simplexmeshIJCV1999}, where each vertex is connected to its two immediate neighbors in the image plane.
These two 1-simplex meshes act as boundary conditions of the surface mesh.
Planar 1-simplex meshes are used to prevent the vertices on the boundaries of the surface mesh from changing their z-values during the deformation;
that is, no contraction or elongation of the entire LV in the direction perpendicular to the SA image planes.
For interjacent vertices, we follow these rules:
(i) vertically, each vertex is connected to its immediate above and below neighbors;
(ii) horizontally, each vertex is connected to its immediate left or right neighbor alternatively, i.e., if one vertex is connected to its immediate left neighbor, then the two vertices next to it should be connected to their right neighbors.
In simplex-mesh geometry, each vertex can only be connected to three neighbors.

\begin{figure}[t]
  \centering
  \includegraphics[width=.49\textwidth]{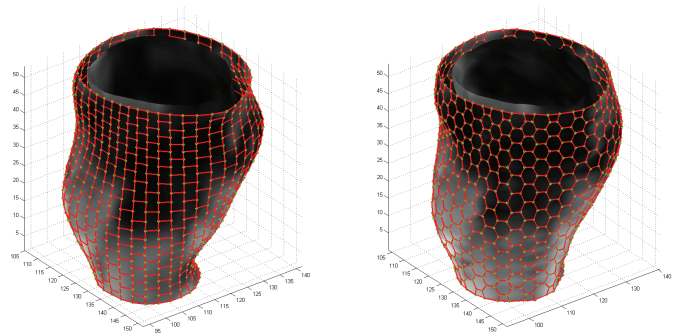}\\
  \caption{Left: illustration of an initial simplex mesh for the epicardial surface; right: illustration of the same simplex mesh after deformation.
  The green dots are the epicardial surface vertices while the red line segments represent the connection relationship among them.
  Also shown in both images is the endocardial surface without its mesh overlaid.}\label{fig:mesh}
\end{figure}

The simplex mesh for the endocardium is initialized in the same way.
One exemplary mesh constructed with the above steps is shown in Fig.~\ref{fig:mesh}.

As in the original simplex mesh work~\citep{simplexmeshIJCV1999}, we define $\bm{F}_{\mathrm{edge}}$ with the notion of closest points.
For each vertex $p_{i}$, we look for the closet pre-detected myocardial edge point $\hat{p}_{i}$ and define the edge attraction force as:
\begin{linenomath}
\begin{equation}\label{eq:F_edge}
    \bm{F}_{\mathrm{edge}} = \omega_{i}\ G \left(\frac{(\hat{p}_{i} - p_{i}) \cdot \bm{n}_{i}}{D_{\mathrm{cutoff}}}\right)\ \bm{n}_{i},\ \mathrm{where}\
    G(x) = \begin{cases}
        x, & \mathrm{if}\ x \leq 1, \\
        0, & \mathrm{if}\ x > 1.
    \end{cases}
\end{equation}
\end{linenomath}
Here, $\omega_{i}$ is a weight between 0 and 1, $\bm{n}_{i}$ the surface normal at vertex $p_{i}$.
The gating function $G(x)$ filters out potential outliers specified by the cutoff distance $D_{\mathrm{cutoff}}$ --
if the projected distance between $\hat{p}_{i}$ and $p_{i}$ onto $\bm{n}_{i}$ is larger than $D_{\mathrm{cutoff}}$, then $\hat{p}_{i}$ is considered an outlier and has no effect on $p_{i}$.
The value of $D_{\mathrm{cutoff}}$ controls the tradeoff between capture range and robustness against outliers during mesh deformation.

\begin{figure}[tp]
  \centering
  \includegraphics[width=.49\textwidth]{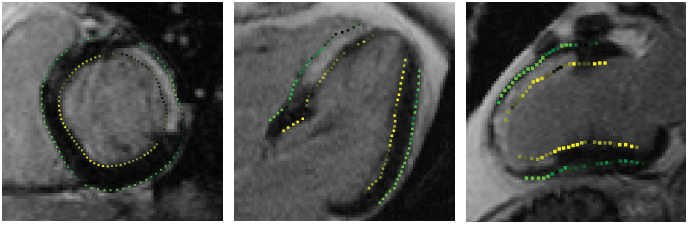}\\
  \caption{The effects of the weights on the detected myocardial edge points: the magnitude of $\omega_{i}$ is coded as brightness of the plot.}\label{fig:weights}
\end{figure}

$\omega_{i}$ is defined as a normalized sum of first-order and second-order absolute intensity differences at $\hat{p}_{i}$.
Recall that $\hat{p}_{i}$ is located on a 1D intensity profile sample $I_{\mathrm{sample}}$.
Assuming it is the $j$th pixel on $I_{\mathrm{sample}}$, we calculate the weighted sum
$S_{i}=|I_{\mathrm{sample}}(j)-I_{\mathrm{sample}}(j-1)| + |I_{\mathrm{sample}}(j)-I_{\mathrm{sample}}(j+1)| + 0.5 * |I_{\mathrm{sample}}(j+1)-I_{\mathrm{sample}}(j-1)|$
and define $\omega_{i}$ by:
\begin{linenomath}
\begin{equation}\label{eq:F_edge_weights}
    \omega_{i} = \frac{S_{i}-\min(\{S_{k}\})}{\max(\{S_{k}\})-\min(\{S_{k}\})},\ \ k=1,2,\cdots,n,
\end{equation}
\end{linenomath}
where $\{S_{k}\}$ is a set of $S_{i}$'s to be normalized together.
For SA slices, we process all the detected endocardial edge points of the entire stack as a set, and all the detected epicardial edge points as another set.
For LA slices, 4C and 2C images are processed separately.
The effects of the weights on the detected myocardial edge points are qualitatively visualized in Fig.~\ref{fig:weights}, where the magnitude of $\omega_{i}$ is coded as brightness of the plot.

In order to define $\bm{F}_{\mathrm{thick}}$, every endocardial vertex is paired with the coplanar epicardial vertex of the same re-sampling angle.
Then a vector-based \emph{spring} force is imposed to maintain the relative position between every pair.
Letting $p_{\mathrm{endo}}^{0}$ denote any endocardial vertex before deformation, and $p_{\mathrm{epi}}^{0}$ its paired epicardial vertex, we define $\bm{F}_{\mathrm{thick}}$ as:
\begin{linenomath}
\begin{equation}\label{eq:F_thick}
    \begin{split}
        &\bm{F}_{\mathrm{epi,thick}}=p_{\mathrm{endo}}^{t}+(p_{\mathrm{epi}}^{0}-p_{\mathrm{endo}}^{0})-p_{\mathrm{epi}}^{t},\\
        &\bm{F}_{\mathrm{endo,thick}}=p_{\mathrm{epi}}^{t}-(p_{\mathrm{epi}}^{0}-p_{\mathrm{endo}}^{0})-p_{\mathrm{endo}}^{t},
    \end{split}
\end{equation}
\end{linenomath}
where $p_{\mathrm{\ast}}^{t}$ is the vertex after $t$th iteration.
$\bm{F}_{\mathrm{thick}}$ is particularly helpful when $\omega_{i}$ is weak at one of the paired vertices but strong at the other.
In this case, as a robust supplement to $\bm{F}_{\mathrm{smooth}}$, $\bm{F}_{\mathrm{thick}}$ extrapolates a reasonable position for the vertex of weak $\omega_{i}$ with the original positional relationship between the pair.

With the meshes initialized and the driving forces defined, we can deform the two surface meshes iteratively with (\ref{eq:ffdScheme})
\footnote{Because the endocardial and epicardial surfaces are generally smooth with small curvatures everywhere, we do not need the advanced adaptation and refinement algorithms in \citep{simplexmeshIJCV1999} and only impose the uniform distribution of vertices via $\bm{F}_{\mathrm{smooth}}$.}.
In each iteration, we first fix the endocardial mesh and update the epicardial mesh, and then update the endocardial mesh with the epicardial mesh fixed.
The deformation is stopped once the movement at any vertex is smaller than 0.1 pixel or the iteration reaches 30 times.
A deformed surface mesh is shown in Fig.~\ref{fig:mesh}, together with its original shape before deformation.
The two meshes obtained after the nonrigid deformation are themselves 3D segmentation of the LV.
To obtain the 2D segmentation of each SA image, we simply intersect it with the final meshes.

\subsection{Implementation}
\label{sec:Method:implementation}
The proposed segmentation framework is implemented under MATLAB R2011a environment, on a PC with Intel Core2 Duo T9400 processer, 3 GB RAM and running 32-bit Microsoft Windows 7 OS.
In practice, we provide the functionality for manual adjustment over the automatic results in each step of the framework.

\section{Experimental results and discussion}
\label{sec:Results}

\subsection{Evaluation on real patient data}
\label{sec:Results:accuracy}

\subsubsection{Data description}
We have evaluated the proposed 3D segmentation framework with 21 sets of real patient data.
The data were acquired with ECG gating by a 1.5T Siemens Symphony MRI scanner from 21 patients (19 males and 2 females, 38-81 years old, mean age $52\pm10$) three months after their having experienced myocardial infarction, following a bolus injection of gadolinium-based contrast agent.
Table~\ref{tab:MRI_parameters} shows the sequence parameters used for the data acquisition.
Cine and LGE sequences of a same subject comprise the same number of SA slices with the same locations.
Depending on the individual heart size, there can be 8-11 SA slices for a patient.
In total there were 206 SA slices for all the 21 patients and 158 slices were selected for infarction analysis according to experts' judgment.
While slices (both SA and LA) of the cine sequence comprise 25 frames, slices of the LGE sequence comprise only one frame.
\begin{table*}[tp]
\fontsize{9pt}{\baselineskip}\selectfont
\centering
\begin{threeparttable}
\caption{Sequence parameters used for the image acquisition.}\label{tab:MRI_parameters}
\begin{tabular}{c|cccccc}
  \hline\hline
              & Flip angle & Repetition / Echo & Width / Height\tnote{a} & Pixel spacing\tnote{b} & Slice thickness / \\
              & [degree$^\circ$] & time [ms] & [pixel] & [mm] & spacing\tnote{c}~~~[mm] \\
  \hline
  LGE images  & 25 & 650$\sim$1000 / 4.18 & 176$\sim$256 & 1.1719$\sim$1.5625 & 7 / 3 \\
  \hline
  Cine images & 55$\sim$74 & 40.88$\sim$44.24 / 1.24$\sim$1.34 & 144$\sim$192 & 1.5625$\sim$1.9792 & 7 / 3 \\
  \hline\hline
\end{tabular}
\begin{tablenotes}
\item[a] Both image width and height are within the same range.
\item[b] Pixel spacing is isotropic in-plane.
\item[c] Slice spacing represents the gap distance between adjacent SA slices.
\end{tablenotes}
\end{threeparttable}
\end{table*}

\begin{figure}[t]
    \centering
    \rotatebox{90}{\hspace{16mm} \footnotesize No. of infarcted instances}
    \begin{minipage}[b]{.45\textwidth}
      \centering
      \includegraphics[width=\linewidth]{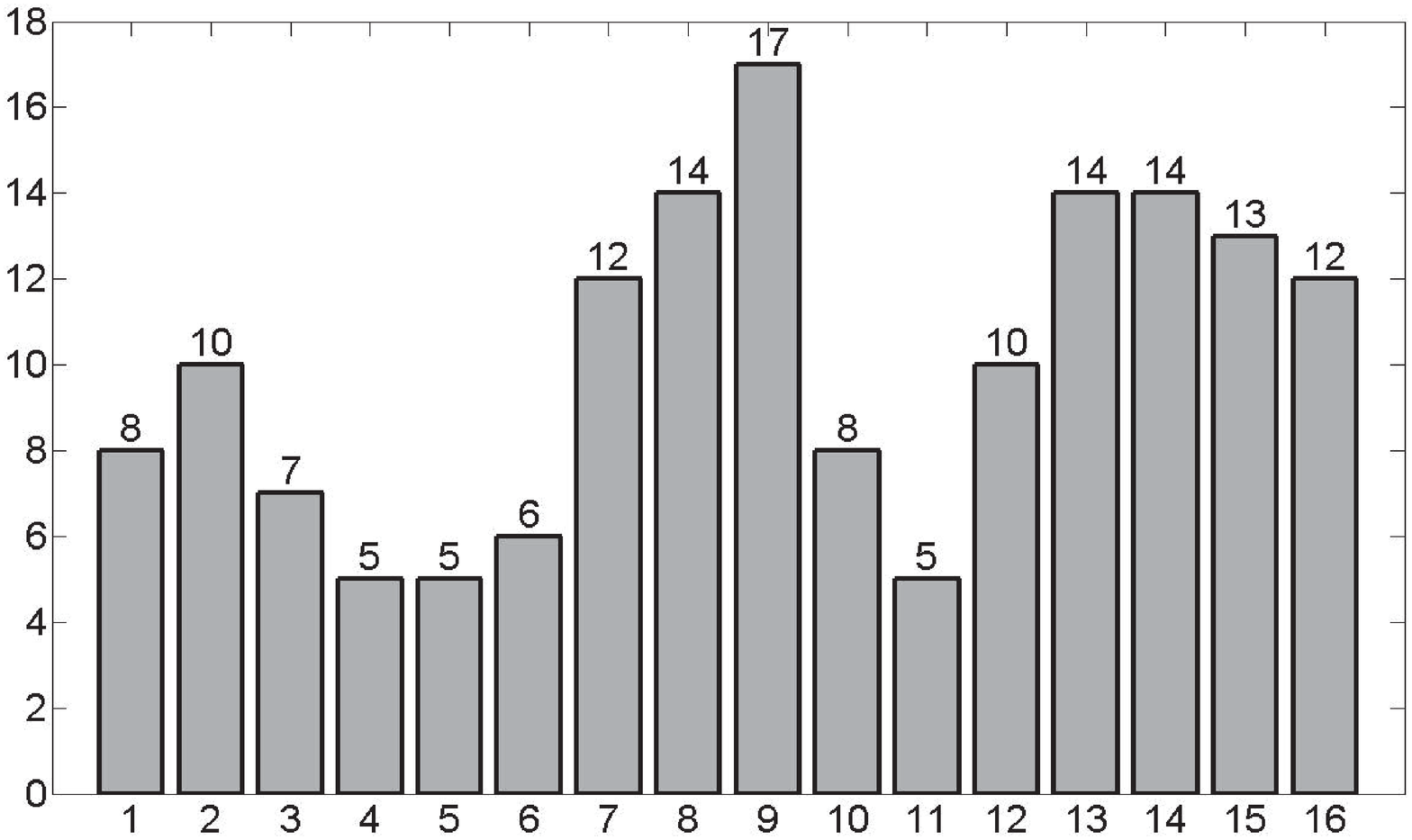}\\
      \footnotesize AHA segment ID.\vspace{3mm}
    \end{minipage}
    \\
    \rotatebox{90}{\hspace{10mm} \footnotesize Average infarct percentage (\%)}
    \begin{minipage}[b]{.45\textwidth}
      \centering
      \includegraphics[width=\linewidth]{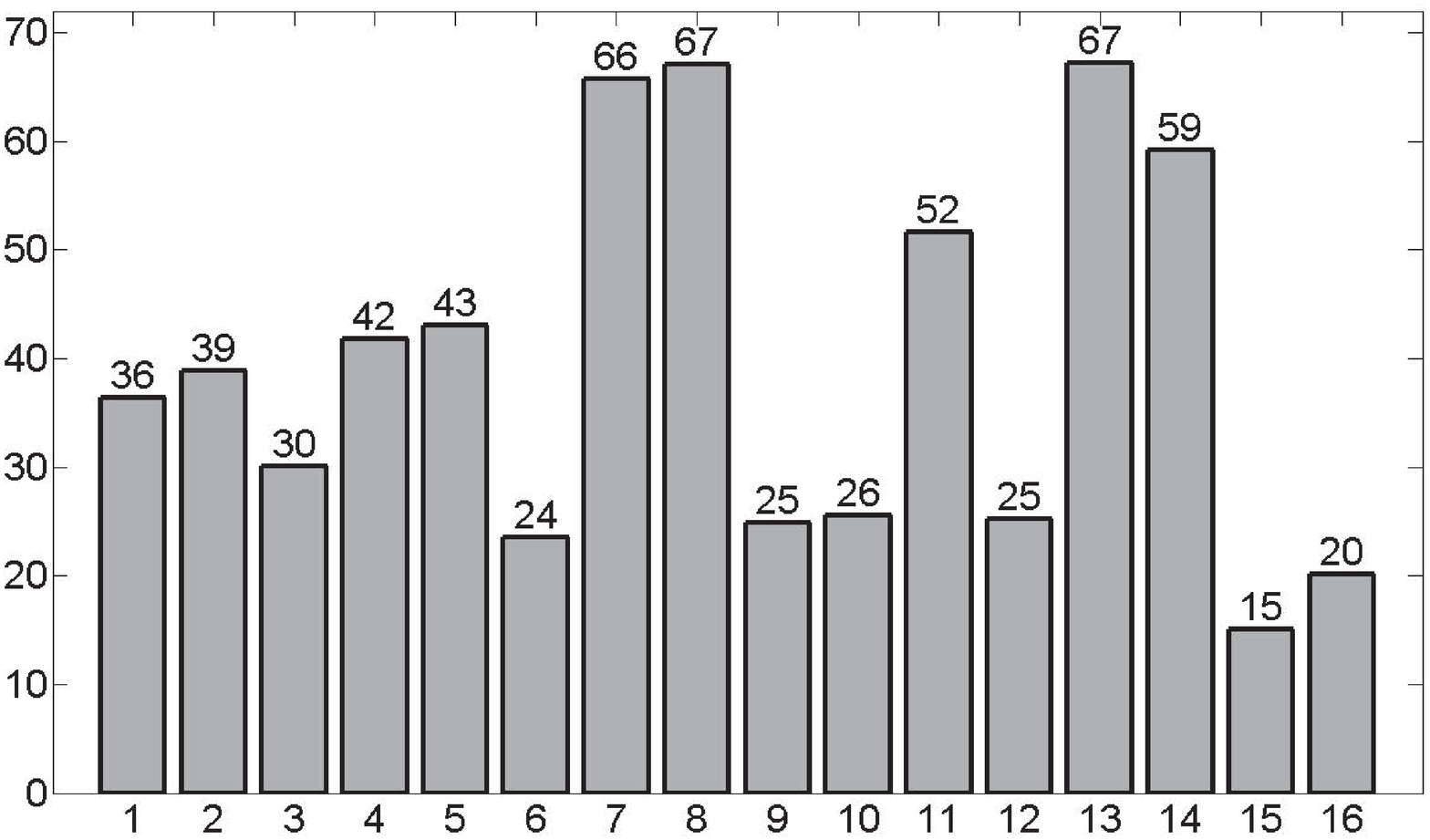}\\
      \footnotesize AHA segment ID.
    \end{minipage}

    \caption{Locality distribution of the infarctions with respect to the AHA 16-segments division.
  Top: number of infarcted instances for each AHA segment (the total number of instances for each segment is 21).
  Bottom: average infarct percentage for each AHA segment, calculated with only infarcted instances.}\label{fig:infarctDistribution}
\end{figure}

The data were manually analyzed (including segmentation of the myocardium and infarcts) by two experts independently, and the manual contours (denoted by $C_{\mathrm{man}1}$ and $C_{\mathrm{man}2}$) were used as the reference standards in our experiments.
The expert who produced $C_{\mathrm{man}1}$ also supervised the production of the \textit{a priori} segmentation in cine data and had full access to the cine data during his analysis of the LGE data, while the other had no access to the cine data.
According to $C_{\mathrm{man}1}$, volumetric percentage of the infarcts with respect to the entire myocardium for the 21 patients ranges from 0 to $35.05\%$ (two cases of non-infarct) with the mean and standard deviation of $18.60\pm 10.43\%$.
Five patients had MVO.
Using the 16-segments division of the LV recommended by the American Heart Association~\citep[AHA;][]{AHA17segments}, 160 out of the total $16\times21=336$ segments (i.e., $47.62\%$) are infarcted, and the locality distribution of the infarctions is charted in Fig.~\ref{fig:infarctDistribution}.

\subsubsection{Experimental settings}
\label{sec:Results:setting}
In the experiments, various parameters introduced in Section~\ref{sec:Method} are set as follows:
(i) For the computation of PI in (\ref{eq:PI}), $r =5$ and $\delta=0.1$.
(ii) $w_{\mathrm{d}}$ and $t_{\mathrm{d}}$ in (\ref{eq:optimalwts}) are confined within a narrow band of 7 pixels for SA images and 9 pixels for LA images centered at $w_{\mathrm{0}}$ and $t_{\mathrm{0}}$.
(iii) For the energy function in (\ref{eq:E_whole}), the weight $\lambda$ is set 0.005 for both SA and LA images.
(iv) For the 3D deformation scheme in (\ref{eq:ffdScheme}), $\gamma=0.7$, $\alpha=0.3$, $\beta=0.3$ and $\mu=0.1$.
(v) For the computation of $\bm{F}_{\mathrm{edge}}$ in (\ref{eq:F_edge}), $D_{\mathrm{cutoff}}=3$.
Although the final meshes produced by the 3D deformation scheme are already a 3D segmentation of the LV, we need to intersect them with the SA images to obtain corresponding 2D segmentation contours (denoted by $C_{\mathrm{auto}}$), in order to compare with manual contours drawn by experts in 2D images.
Minimum manual adjustments were performed to better evaluate the proposed framework.
Of all the 158 SA slices selected for analysis, 3 slices (each from a different volume) were manually reallocated after the automatic misalignment correction, and 8 slices were manually registered due to failure of the automatic translational registration.
No manual adjustments were performed on the final $C_{\mathrm{auto}}$.

\subsubsection{Results}
\begin{figure*}[t]
  \centering
  \includegraphics[width=\textwidth]{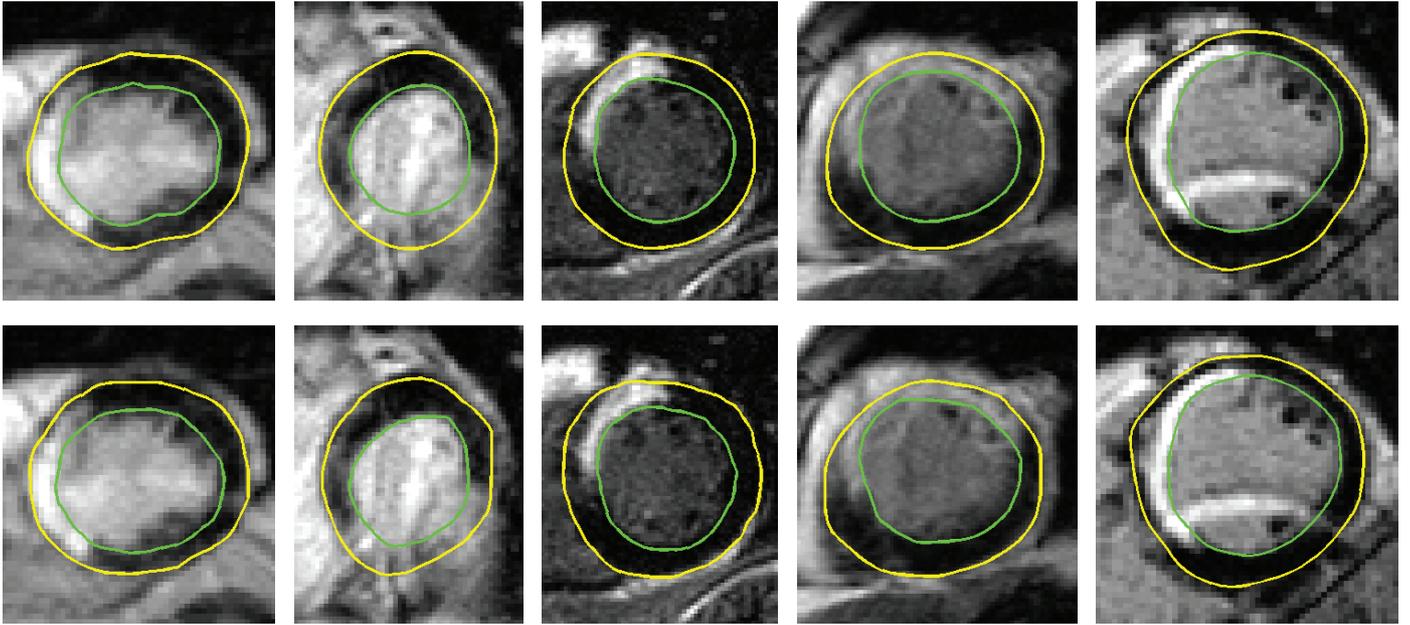}\\
  \caption{Some exemplary segmentation results of our automatic framework ($C_{\mathrm{auto}}$, top row), as compared to those by one of the experts ($C_{\mathrm{man}1}$, bottom row).}\label{fig:auto-manual_results}
\end{figure*}

Qualitatively, we observe that segmentation results produced by our framework are consistently accurate for nearly all the LGE images in our database.
Figure~\ref{fig:auto-manual_results} shows some exemplary results, together with the manual contours delineated by one of the experts.
Quantitatively, we have evaluated our framework in terms of both distance- and region-based measures.
We have calculated mean distance error and the Dice coefficient \citep{Dice1945} between the automatic results and reference standards slice by slice.
We have also calculated volumetric Dice coefficients.
The results are presented in Table \ref{tab:accurcy}.
Of the three groups of comparisons, $C_\mathrm{auto}$ and $C_{\mathrm{man}1}$ are the most similar with extremely small distances and high slice-wise and volumetric Dice coefficients.
Although the difference between $C_\mathrm{auto}$ and $C_{\mathrm{man}2}$ is slightly larger, it is very close to the reported inter-observer variations between $C_{\mathrm{man}1}$ and $C_{\mathrm{man}2}$.
These results confirm that the myocardial segmentations produced by our framework are close to those obtained by the experts.
The general trend that the difference between $C_\mathrm{auto}$ and $C_{\mathrm{man}1}$ is smaller than between $C_\mathrm{auto}$ and $C_{\mathrm{man}2}$ is not surprising, because the expert who produced $C_{\mathrm{man}1}$ also supervised the \textit{a priori} segmentation of the cine data.
On the contrary, the other expert had no such \textit{a priori} knowledge and had to guess based on her experience when the contrast is poor.
\begin{table*}
\fontsize{9pt}{\baselineskip}\selectfont
  \centering
  \caption{The segmentation accuracy evaluated from the aspects of the mean distance errors and Dice coefficients.}\label{tab:accurcy}
  \begin{tabular}{l|r|ccc}
    \hline\hline
     &  & $C_{\mathrm{man}1}$ vs. $C_{\mathrm{man}2}$ & $C_\mathrm{auto}$ vs. $C_{\mathrm{man}1}$ & $C_\mathrm{auto}$ vs. $C_{\mathrm{man}2}$ \\
     \hline
                              & Endocardium & 1.46$\pm$0.66 & 0.94$\pm$0.44 & 1.51$\pm$0.74 \\
     Mean distance error [mm] & Epicardium  & 1.56$\pm$0.58 & 0.90$\pm$0.41 & 1.68$\pm$0.70 \\
                              & Both        & 1.46$\pm$0.47 & 0.90$\pm$0.36 & 1.54$\pm$0.56 \\
     \hline
                                      & LV         & 94.76$\pm$2.25 & 96.88$\pm$1.84 & 94.35$\pm$2.70 \\
     Slice-wise Dice coefficient [\%] & BP         & 92.77$\pm$4.29 & 95.33$\pm$3.62 & 92.64$\pm$4.36 \\
                                      & Myocardium & 82.93$\pm$5.28 & 88.57$\pm$4.75 & 82.32$\pm$5.59 \\
     \hline
                                      & LV         & 95.06$\pm$0.91 & 97.18$\pm$0.48 & 94.78$\pm$0.88 \\
     Volumetric Dice coefficient [\%] & BP         & 93.65$\pm$1.47 & 95.98$\pm$0.87 & 93.42$\pm$1.96 \\
                                      & Myocardium & 83.49$\pm$2.49 & 89.19$\pm$1.42 & 82.84$\pm$2.50 \\
    \hline\hline
  \end{tabular}
\end{table*}

\subsubsection{Comparison with related works}
It is difficult to quantitatively compare the accuracy of our 3D segmentation framework with those of the existing methods~\citep{relatedworks01_Ciofolo2008,relatedworks02_Dikici2004}.
Given that different data sets were used, it is inappropriate to directly compare accuracies or errors reported in the papers.
However, qualitatively we have identified the following advantages of our 3D segmentation method over others.

\begin{figure}[tp]
    \centering
    \includegraphics[width=.49\textwidth]{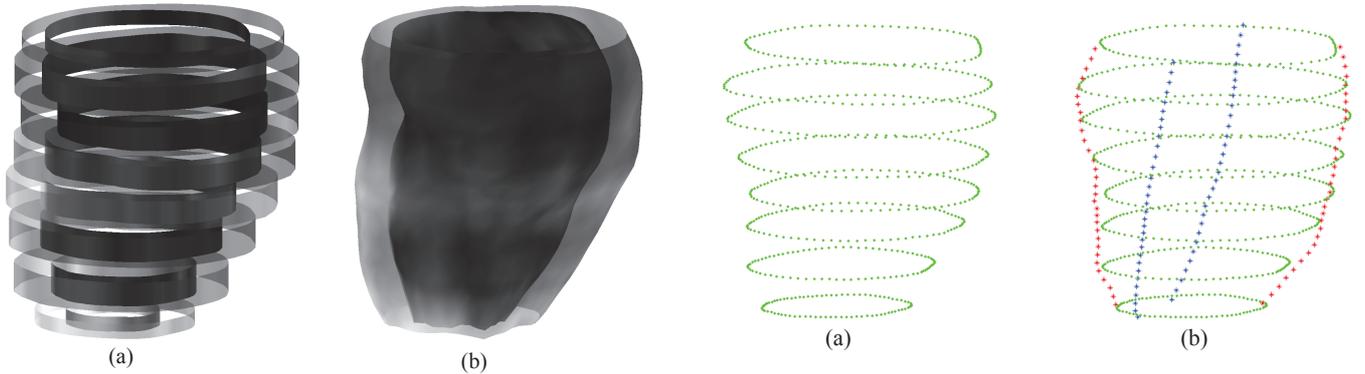}\\
    \caption{Qualitative comparison of 2D and 3D segmentation:
    (a) Pure 2D segmentation methods produce discrete cylinders in 3D space;
    (b) Our 3D segmentation achieves more accurate 3D reconstruction of the LV.
    Here epicardial surfaces are made transparent for visualization.}\label{fig:2d3dComparison}
\end{figure}

First, compared to pure 2D segmentation methods \citep{relatedworks02_Dikici2004,myownwork} which merely produce discrete cylinders in 3D space (Fig.~\ref{fig:2d3dComparison}(a)), our 3D segmentation framework produces more accurate 3D reconstruction of the LV geometry (Fig.~\ref{fig:2d3dComparison}(b)).
Moreover, the 3D smoothness force $\bm{F}_{\mathrm{smooth}}$ helps extrapolate reasonable myocardial boundaries with information from the LA and neighboring SA slices, when the current slice lacks contrast.

\begin{figure}[tp]
  \centering
  \includegraphics[width=.475\textwidth]{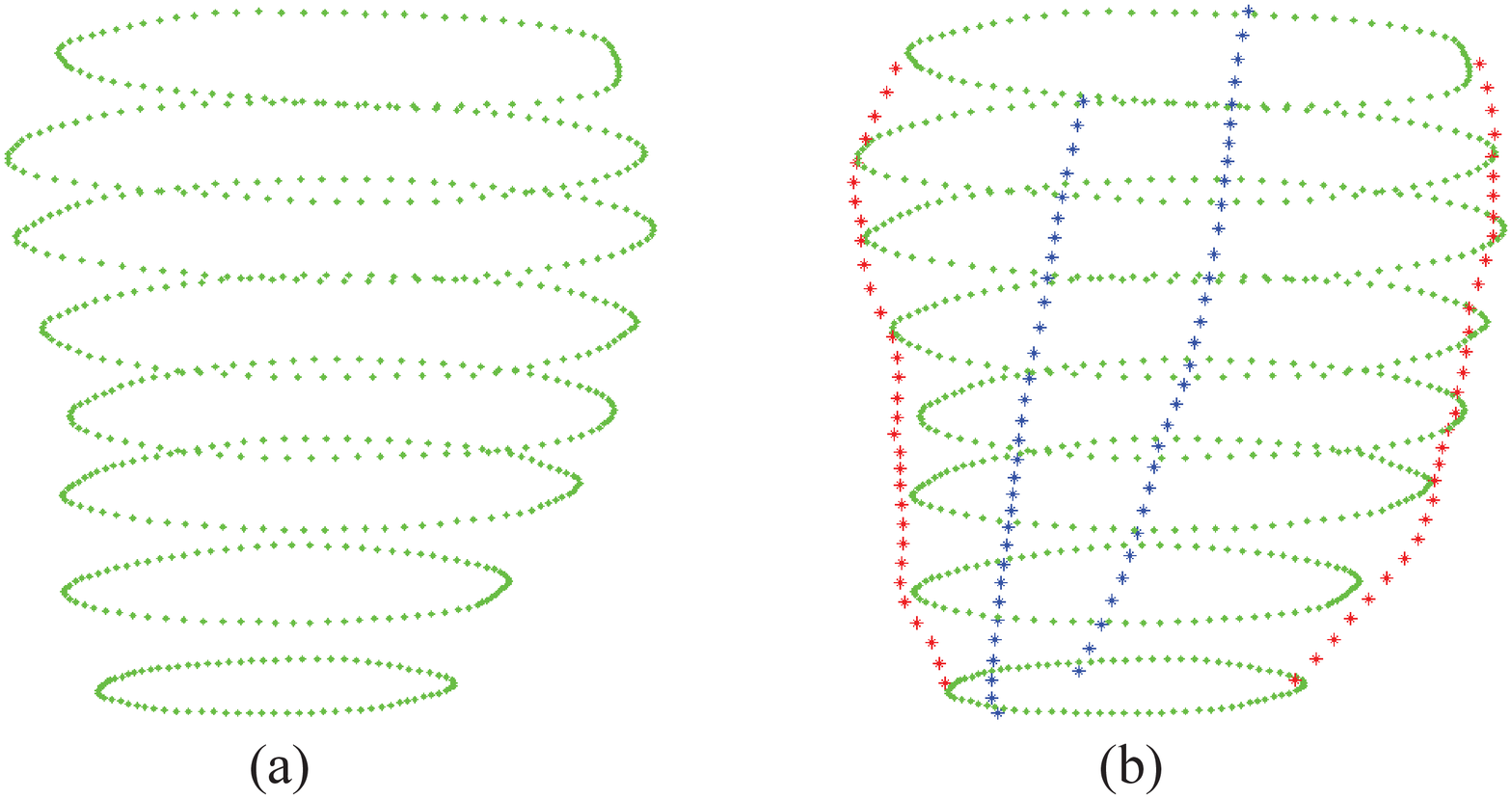}\\
  \caption{Detected epicardial edge points displayed in 3D: (a) only the SA images are used for the detection; (b) edge points from 4C (blue) and 2C (red) LA images are added, providing a considerable amount of extra information between the SA images.}\label{fig:3dFeaturePts}
\end{figure}

Second, though the work by \cite{relatedworks01_Ciofolo2008} also involved deforming 3D meshes for segmentation, the meshes were attracted only to features detected in SA LGE images.
Due to the large anisotropy of the stacks of SA images, a respectable portion of the vertices was moved by only regulating forces because there were no feature points lying between the SA slices (Fig.~\ref{fig:3dFeaturePts}~(a)).
In contrast, since we incorporate the two standard LA views (4C and 2C) into our 3D segmentation framework, a considerable amount of feature points detected in the LA images fill the gaps (Fig.~\ref{fig:3dFeaturePts}~(b)).
These extra feature points can make the final meshes represent the physical shape of the scanned LV with higher fidelity.
The 2D segmentation contours in SA images obtained by intersection with the meshes can also be more accurate, because the extra information contained between the SA images is propagated over the meshes by $\bm{F}_{\mathrm{smooth}}$ \citep{simplexmeshIJCV1999}.

Third, the work by \cite{relatedworks01_Ciofolo2008} did not mention any correction of the misalignment artifacts for stacks of LGE images.
This exposes the reconstructed 3D representation of the LV to potential distortion~\citep{heartMotionStudy}.
In contrast, we realign all the SA and LA slices in the original patient coordinate system prior to the construction and deformation of the meshes, thus endowing the 3D segmentation with higher fidelity.


\subsection{Evaluation on simulated data}
\subsubsection{Data generation and experimental settings}
\begin{figure}[tp]
    \centering
    \includegraphics[width=.49\textwidth]{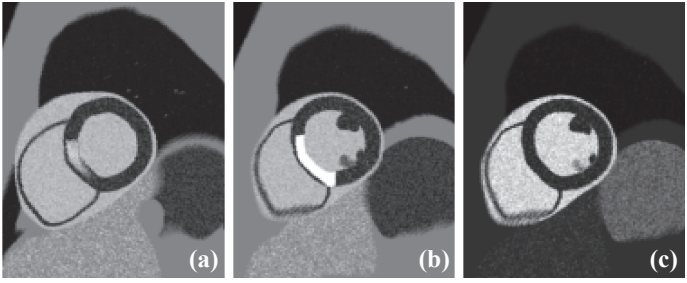}
    \caption{Examples of the simulated data: (a)-(b) LGE and (c) cine images.}\label{fig:phantomExamples}
\end{figure}

We also evaluated our framework on simulated data generated from the 4D XCAT male phantom \citep{XCAT}.
The LGE data are simulated as follows.
First a voxelized torso phantom with an isotropic voxel size of 1.34 mm is generated from the XCAT.
Nine tissues are included: the myocardium, infarct, blood, pericardium, lung, spleen, liver, stomach and the rest of the body.
An expert manually delineated regions for each tissue in a set of real patient data.
Then every voxel in the torso phantom is randomly assigned a value according to its tissue type.
The assigned value is uniformly sampled from the regions of the specified tissue.
Two exceptions are the infarct and body: voxels of these two tissues are assigned mean values of the respective regions.
After that, simulated images of standard CMR orientations (SA, 4C and 2C) with the same slice dimensions as our real data (pixel spacing = 1.34 mm, slice thickness and spacing = 7 and 3 mm, resulting in 8 SA slices) are reconstructed from the torso volume.
Each simulated image is an average of the intensities within its slice thickness to increase the signal to noise ratio.
The cine data are simulated in the same way, except that no infarct is included.
Figure~\ref{fig:phantomExamples} shows several examples of the simulated cine and LGE data.

To create enough difference in myocardium shape between the LGE and cine SA images, the LGE and cine data were simulated at different cardiac and respiratory phases.
The infarcts were set at mid-ventricle, spanning 50 mm along the LA of the LV.
Four LV volumes with infarcts located at anterior, lateral, inferior and septal walls respectively were simulated.
The infarcts span 100 degrees on the myocardial circumference and are 100\% transmural.
Myocardial contours of the simulated data were derived from the original voxelized phantom.
The cine contours are used as the \textit{a priori} segmentation to our segmentation framework, while the LGE contours are used as the ground truth ($C_{\mathrm{grnd}}$) for evaluation.

The various parameters are set to the same as presented in Section \ref{sec:Results:setting}.
No manual adjustment was performed to the simulated data.

\subsubsection{Results}
The segmentation results obtained by an expert and our framework on the simulated LGE data are shown in Table \ref{tab:accurcyOnPhantom}.
Both manual and automatic segmentations are very close to the ground truth, and the difference between the two segmentations is also small.
What is noteworthy is the standard deviation of the slice-wise Dice coefficients for the BP between $C_\mathrm{auto}$ and $C_{\mathrm{grnd}}$, which is significantly higher than the others.
This is caused by the most apical slice in each volume and can be explained by three facts regarding very apical slices:
(i) the myocardium is usually much blurred;
(ii) the apex's motion is large and causes a big change between the myocardium shape in cine and LGE images;
(iii) areas of the myocardium are small and the Dice coefficient is known to be sensitive to small areas.
In fact segmentation of the myocardium in very apical slices is always so difficult that no method can consistently produce reliable results.
In practice, we provide a simple user correction scheme to effectively handle such cases.
The mean Dice coefficient for the BP between $C_\mathrm{auto}$ and $C_{\mathrm{grnd}}$ becomes $96.97\pm1.50\%$ if the most apical slice in each volume is excluded.
\begin{table*}
\fontsize{9pt}{\baselineskip}\selectfont
  \centering
  \caption{The segmentation accuracy evaluated with the simulated data.}\label{tab:accurcyOnPhantom}
  \begin{tabular}{l|r|ccc}
    \hline\hline
     &  & $C_{\mathrm{man1}}$ vs. $C_{\mathrm{grnd}}$ & $C_\mathrm{auto}$ vs. $C_{\mathrm{grnd}}$ & $C_\mathrm{auto}$ vs. $C_{\mathrm{man1}}$ \\
     \hline
                              & Endocardium & 0.68$\pm$0.17 & 0.73$\pm$0.49 & 0.71$\pm$0.34 \\
     Mean distance error [mm] & Epicardium  & 0.81$\pm$0.42 & 0.67$\pm$0.41 & 1.12$\pm$1.04 \\
                              & Both        & 0.76$\pm$0.28 & 0.69$\pm$0.44 & 0.92$\pm$0.67 \\
     \hline
                                      & LV         & 95.78$\pm$4.70 & 96.93$\pm$3.74  & 94.22$\pm$8.42 \\
     Slice-wise Dice coefficient [\%] & BP         & 94.81$\pm$4.35 & 92.86$\pm$11.20 & 93.79$\pm$8.32 \\
                                      & Myocardium & 90.86$\pm$5.30 & 92.73$\pm$5.51  & 88.61$\pm$11.13 \\
     \hline
                                      & LV         & 97.44$\pm$0.14 & 98.05$\pm$0.07 & 96.99$\pm$0.18 \\
     Volumetric Dice coefficient [\%] & BP         & 96.66$\pm$0.27 & 96.99$\pm$0.19 & 96.81$\pm$0.02 \\
                                      & Myocardium & 92.58$\pm$0.50 & 94.09$\pm$0.17 & 91.80$\pm$0.42 \\
    \hline\hline
  \end{tabular}
\end{table*}

%

\subsection{Robustness with respect to different a priori segmentations}
In order to test the consistency of the 3D segmentation framework with different \emph{a priori} segmentations, we have conducted two experiments with both practical and artificially simulated \emph{a priori} segmentations, respectively.
We select a set of LGE data with fair segmentation accuracy for the experiments, since our focus here is the consistency instead of accuracy.

\subsubsection{Experiment 1: segmentation with different practical a priori segmentations}
In this experiment, one observer manually drew the myocardial contours for the selected SA cine images that correspond to the target LGE images, serving as the manual \emph{a priori} segmentation (denoted by $A_{\mathrm{manu}}$).
Meanwhile, the automatically generated cine contours used in Section~\ref{sec:Results:accuracy} (denoted by $A_{\textrm{auto}}$) serve as the control group.
Segmentation accuracies with $A_{\mathrm{manu}}$ and $A_{\mathrm{auto}}$ as \emph{a priori} are compared to test whether the segmentation framework can produce consistently good results with both manually drawn and automatically generated \emph{a priori} segmentations.
The results are presented in Table~\ref{tab:perturbedApriori}.
As we can see, the segmentation results with both kinds of inputs are similarly good, though the results produced with $A_{\mathrm{manu}}$ are slightly better.
This indicates that our framework can consistently produce high quality segmentation for LGE images despite that in practice the potential \emph{a priori} segmentation in the corresponding cine images can be either manually drawn or (semi-) automatically generated.
\begin{table*}
\fontsize{9pt}{\baselineskip}\selectfont
  \centering
  \caption{Segmentation accuracies with different practical \emph{a priori} segmentations. The reference standard used here was $C_\mathrm{man1}$. }\label{tab:perturbedApriori}
  \begin{tabular}{r|ccc|ccc}
    \hline\hline
     & \multicolumn{3}{c|}{Mean distance error [mm]} & \multicolumn{3}{c}{Dice coefficient [\%]} \\
    \emph{A priori} & Endocardium & Epicardium & Both & LV & BP & Myocardium \\
    \hline
    $A_{\mathrm{auto}}$ & 1.04$\pm$0.44 & 1.19$\pm$0.32 & 1.12$\pm$0.29 & 96.05$\pm$1.08 & 94.89$\pm$3.33 & 86.18$\pm$3.76 \\
    $A_{\mathrm{manu}}$ & 0.96$\pm$0.33 & 1.08$\pm$0.26 & 1.02$\pm$0.18 & 96.53$\pm$0.59 & 95.65$\pm$1.91 & 87.84$\pm$2.73 \\
    \hline\hline
  \end{tabular}
\end{table*}

\subsubsection{Experiment 2: segmentation with different simulated a priori segmentations}
In this experiment, $A_{\mathrm{manu}}$ is purposely enlarged and shrunk by 1 mm to generate two sets of artificially simulated \emph{a priori} segmentations -- $A_{\mathrm{enlg}}$ and $A_{\mathrm{shrk}}$.
Differences between segmentation results with $A_{\mathrm{enlg}}$ and $A_{\mathrm{shrk}}$ as \emph{a priori} segmentations are examined to test the consistency of our segmentation framework given the \emph{a priori} segmentations 2 mm apart from each other.
The comparison is presented in Table~\ref{tab:SegmentApriori}.
Mean distances between the segmentation results are greatly decreased from the mean distances between the two \emph{a priori} themselves.
Area similarity is also improved, which is revealed by the fact that Dice coefficients evaluated at all three levels are significantly increased.
The comparison results further validate the consistency of our segmentation framework -- it can produce consistent segmentations even with varied \emph{a priori} in the range of 2 mm.
\begin{table*}[htbp]
\fontsize{9pt}{\baselineskip}\selectfont
  \centering
  \caption{Comparisons between $A_{\mathrm{enlg}}$ and $A_{\mathrm{shrk}}$, and between segmentation results with $A_{\mathrm{enlg}}$ and $A_{\mathrm{shrk}}$ as \emph{a priori} segmentations.}\label{tab:SegmentApriori}
  \begin{tabular}{r|ccc|ccc}
    \hline\hline
     & \multicolumn{3}{c|}{Mean distance [mm]} & \multicolumn{3}{c}{Dice coefficient [\%]} \\
    Comparison between & Endocardium & Epicardium & Both & LV & BP & Myocardium \\
    \hline
    $A_{\mathrm{enlg}}$ and $A_{\mathrm{shrk}}$ & 1.96$\pm$0.03 & 1.98$\pm$0.01 & 1.97$\pm$0.02 & 93.44$\pm$1.23 & 90.59$\pm$2.66 & 77.24$\pm$2.96 \\
    \hline
    Segmentation results & 1.21$\pm$0.21 & 1.12$\pm$0.24 & 1.16$\pm$0.22 & 96.29$\pm$1.27 & 94.46$\pm$1.43 & 87.03$\pm$4.03 \\
    \hline\hline
  \end{tabular}
\end{table*}

\subsubsection{Summary}
The experimental results have shown that our framework can produce highly consistent segmentations despite that the given \emph{a priori} segmentation in cine images can vary.
The consistency and robustness are important and desirable because both inter-observer variability and automatic segmentation accuracies for cine images are in the range of 1-2 mm \citep{cineSegmentationReview2011}.
Although starting with propagating the \emph{a priori} segmentation contours from cine images, the consistency of our segmentation framework makes it possible to obtain highly reproducible segmentation of LGE images given different \emph{a priori} segmentations (either manual or automatic) in cine images.

\subsection{Study limitations}
The proposed framework was evaluated with a small number of patients with chronic infarction.
In acute infarction, the pathology and signal enhancement pattern can vary drastically and performance of the framework is uncertain if applied to such cases.
To test full clinical capacity of the framework, future experiments with a much larger cohort including both acute and chronic infarction patients are needed.
Another limitation is that we did not use independent training and testing datasets in this work.
Although the framework is not learning based and found to be insensitive to small fluctuations of the parameters, it is still necessary to test the framework's clinical usability with independent training and testing sets in the future.

We did not explicitly handle cases of signal intensity bias in the myocardium, such as imperfect inversion time, surface coil intensity drop off, or the existence of MR imaging artifacts.
Such cases within a reasonable range are implicitly countered by the robustness of the framework.
For example, the last column of Figure~\ref{fig:auto-manual_results} shows an LGE slice in our database with a significant bright artifact and our framework still produced correct myocardial contours.
However, severe cases that even cause troubles to human observers are considered imaging failures and beyond our scope.
Lastly, we did not consider enhancements in the LGE images due to other pathologies, which may present complex enhancement patterns that are not covered by the proposed model.
But the focus on a single pathology (i.e., ischemic heart disease) makes the framework more reliable for the target subject group.

\section{Conclusion}
\label{sec:Conclusion}
This paper presents a novel 3D segmentation framework of the LV in LGE CMR images.
Unlike most related works \citep{relatedworks02_Dikici2004, relatedworks01_Ciofolo2008, myownwork} which merely used SA images, we also incorporate the standard 4C and 2C LA images to provide supplementary information in the big gaps between contiguous SA slices.
Compared to another segmentation method~\citep{relatedworks01_Ciofolo2008} which also involved 3D mesh deformation, we realign all the SA slices in a unified 3D coordinate system prior to the mesh construction and deformation, to eliminate potential distortion of the 3D representation caused by misalignment artifacts.
In addition, we propose a novel parametric model of the LV in LGE images based on 1D intensity profiles.
This model is able to simultaneously detect paired endocardial and epicardial edge points with varied myocardium thickness, adaptively distinguish between healthy and infarcted myocardium, and readily be applied to both SA and LA images.
It is embedded into an energy minimization scheme for reliable detection of myocardial edge points.
The experimental results on both real patient and simulated phantom data have shown that our framework is able to generate accurate segmentation for LGE images and is robust with respect to varied \emph{a priori} segmentation in the referenced cine images.

\section*{Acknowledgements}
The authors would like to thank the Academic Research Fund, National University of Singapore, Ministry of Education, Singapore for funding the CMR studies.
We are also grateful to the radiographers and staff at the Department of Diagnostic Imaging, National University Hospital, Singapore, for helping with the CMR scans.

\bibliographystyle{model2-names}

\end{document}